%% file: ms.tex
\documentclass[12pt,preprint]{aastex}

\begin{document}

\title{Multiwavelength Observations of the Extreme X-Ray Selected BL Lac Object PG 1553+11 (1ES 1553+113)}

\shorttitle{PG 1553+11; an Extreme XBL}
\shortauthors{Osterman et al.}

\author{M. Angela Osterman\altaffilmark{a}, H. Richard Miller\altaffilmark{a}, Amy M. Campbell\altaffilmark{a,c}, Kevin Marshall\altaffilmark{a}, John P. McFarland\altaffilmark{a,h}, Hugh Aller\altaffilmark{b}, Margo Aller\altaffilmark{b}, Robert E. Fried\altaffilmark{d}\footnote{Passed away on November 13, 2003}, Omar M. Kurtanidze\altaffilmark{e,f,g}, Maria G. Nikolashvili\altaffilmark{e}, Merja Tornikoski\altaffilmark{i}, and Esko Valtaoja\altaffilmark{j}}

\altaffiltext{a}{Dept. of Physics \& Astronomy, Georgia State University, 1 Park Place ste. 730, Atlanta, GA, 30303; osterman@chara.gsu.edu}

\altaffiltext{b}{University of Michigan, Dept. of Astronomy, 500 Church St. 830 Dennison, Ann Arbor, MI, 48109-1042}

\altaffiltext{c}{Louisiana State University, Dept. of Physics \& Astronomy, 202 Nicholson Hall, Tower Dr., Baton Rouge, LA 70803-4001}

\altaffiltext{d}{Braeside Observatory, 8055 W. Naval Observatory Rd., Flagstaff, AZ, 86001}

\altaffiltext{e}{Abastumani Observatory, Georgian Academy of Sciences, Mt Kanobil, 383762, Georgia}

\altaffiltext{f}{Landessternwarte Heidelberg, D-69117 Heidelberg, Germany}

\altaffiltext{g}{Astrophysikalisches Institut Potsdam, An der Sternwarte 16, D-14482 Potsdam Germany}

\altaffiltext{h}{Kapteyn Instituut, Rijksuniversiteit Groningen, 9747 AD Groningen, The Netherlands}

\altaffiltext{i}{Mets\"{a}hovi Radio Observatory, Mets\"{a}hovintie 114, FIN-02540 Kylm\"{a}l\"{a}, Finland}

\altaffiltext{j}{Tuorla Observatory, FIN--21500 Piikki\"{o}, Finland}

\begin{abstract}
PG 1553+11 was the target of a coordinated three week multiwavelength campaign during 2003 April and May.
A significant X-ray flare was observed during the second half of this campaign.
Although no optical flare was recorded during the X-ray campaign, optical observations obtained immediately prior to the campaign displayed a higher flux than that recorded during the campaign. 
An optical flare was observed a few days after the end of the X-ray campaign and may be related to the X-ray flare.
Radio observations were made at three frequencies, with no significant changes in flux detected near the times of the optical and X-ray flares.
The spectral energy distributions and flux ratios in different wavebands observed for this object are compared to other X-ray selected blazars to demonstrate how PG 1553+11 is an extreme member of this group. 
\end{abstract}


\section{Introduction}

In the broad class of extragalactic objects known as Active Galactic Nuclei (AGN), blazars distinguish themselves in many ways. 
Blazars are a radio-loud subclass of AGN, typically featuring very core-dominated radio morphologies. 
Their optical continua are featureless and markedly steep, they exhibit a high degree of polarization (up to about 20\%), and their fluxes vary on timescales of an hour to several years at all observed wavelengths.
The most extreme and unique property of blazars is their highly beamed continuum, most likely produced by a jet of relativistic material aimed close to the observer's line of sight. 
BL Lac objects have particularly weak spectral lines, so their redshifts, and hence luminosities, are difficult to determine.
See \citet{urr95} and references therein for further details.

The spectral energy distribution (SED) of blazars is distinguished by two peaks: one in the radio/UV regime and the other in the X-ray/$\gamma$-ray regime. 
The spectrum in the radio to UV range is generally agreed to arise from synchrotron emission from relativistic electrons spiralling around the jet's magnetic field lines \citep{har96}. 
The X-ray/$\gamma$-ray spectrum is most likely due to inverse Compton (IC) emission in radio selected blazars (RBLs), while most of the X-rays are the high energy tail of the synchrotron emission in X-ray selected blazars (XBLs) \citep{urr95}. 

A major unsolved puzzle of blazars is what supplies the ``seed photons" which are upscattered to produce the IC emission. 
The X-ray variability of blazars is observed to be more pronounced than that of longer wavelengths, implying that the source of the X-rays is located close to the central engine. 
In order to fully understand blazars, we must understand the physics of the region near the central engine, since this is where jet particles are collimated and accelerated to relativistic speeds. 

PG 1553+11 was identified in the Palomar-Green survey of ultraviolet-excess objects with a 15.5 magnitude blue stellar object \citep{gre86}. 
\citet{mil88} identified this object as a blazar with $z = 0.36$ and an optical R magnitude varying from $\sim$13 to $\sim$15.5.
Subsequent observations revealed that this object has rather weak radio emission for a blazar, exhibiting fluxes of a few tenths of a Jy at 5.0 GHz \citep{fal90}. 
Its SED indicates that it is an XBL \citep{fal90,don05}.
Its X-ray properties were studied in the Einstein Slew Survey of BL Lac objects, and were included in the catalog as 1ES 1553+113 \citep{per96}.
The relatively weak radio emission suggests that this is an extreme XBL. 
There is only one previous pointed RXTE observation of this object, it was detected at a flux of $3.8\times10^{-11}\mbox{ erg cm}^{-2}\mbox{ sec}^{-1}$.
The RXTE archives also include some All Sky Monitor pointings to this object. 

\section{Observations and Data Reduction}

The X-ray observations began as the result of a Target of Opportunity (ToO) RXTE program triggered by an optical flare.
The observations were recorded using RXTE's Proportional Counter Array (PCA).
Higher frequency (37 GHz) radio observations were obtained with the Mets\"{a}hovi Radio Observatory's 13.7m dish.
Lower frequency (14.5 and 4.8 GHz) radio observations were obtained with the 26m dish at the University of Michigan Radio Astronomy Observatory (UMRAO).
Optical (R and B band) observations were obtained with the 0.7m reflector at the Abastumani Observatory in Georgia, the Perkins 72" (1.8m) reflector at Lowell Observatory in Flagstaff, Arizona, and the 16" (0.4m) reflector at Braeside Observatory in Arizona.
All optical observations were obtained using CCDs and through standard BVRI filters.

The RXTE observations span about three weeks in Spring of 2003; April 22$ - $May 12.
Observations with integration times of 3 ksec were performed $2-3$ times per day for 21 days.
Optical observations were obtained regularly during most of the RXTE campaign, as well as in April, late May, and July.
Radio observations at 14.5 and 37.0 GHz were obtained regularly during most of the RXTE campaign, along with one 4.8 GHz observation.
The 14.5 and 4.8 GHz observations continued through early September.
A summary of all observations obtained during this campaign is shown in Table 1.
Figures 1$ - $3 display the lightcurves of all data published here. 

\subsection{X-ray Data Reduction}

The X-ray light curve was extracted using the {\sc FTOOLS}~v5.2 software package.
During nearly all of our observations, PCUs 1, 3, and 4 were turned off.
Therefore, data were only extracted from PCUs 0 and 2.
Despite the loss of the propane layer onboard PCU 0 during May of 2000, the signal to noise ratio was much greater when using data from both PCUs 0 and 2.
To further enhance the signal-to-noise ratio, only data from layer 1 of the PCA were analyzed.
No data from the HEXTE cluster or the other PCUs were used.
All of the data analyzed here were taken while the spacecraft was in {\sc STANDARD-2} data mode.

Data were extracted only when the target's Earth elevation angle was $>10\deg$, pointing offset $<0.02\deg$, PCUs 0 and 2 both on, the spacecraft more than 30 minutes after SAA passage, and electron noise less than 0.1 units.
Since the background response of the PCU is not well defined above 20~keV, only channels $0-44$ ($2-20$~keV) are included in this analysis.

Because the PCA is a non-imaging detector, background issues can be critically important during analysis.
The faint-mode ``L7'' model, developed by the PCA team, was used here.
This model provides adequate background estimation for objects with less than 40 cnts/sec.
Background files were extracted using {\tt pcabackest}~v3.0.

\subsection{Results of X-ray Observations}

Observations utilizing integrations of about 3 ksec were performed 2-3 times per day with the PCA for 21 days. 
The RXTE campaign ran from 2003 April 21 to May 12, consisting of a total of 46 X-ray observations. 
Table 2 features all of the light curve and spectral slope data taken for this campaign. 
The dense, regularly sampled X-ray data display a substantial, well-defined flare starting almost halfway through the campaign, as displayed in Figures 1 and 2.
These figures also display the simultaneous lightcurves of the optical and radio data for comparison.

The X-ray flux more than doubled over a period of ten days, and, although the observations stopped before the flare was complete, they indicated that the flare had already reached its peak and was beginning to return to lower flux levels. 
The flare began on April $29-30$ and peaked on May 10. 
The observations record an average of about $5.0\times10^{-12}\mbox{ erg cm}^{-2}\mbox{ sec}^{-1}$ before the flare began, while at the flare's peak the flux approached $1.2\times10^{-11}\mbox{ erg cm}^{-2}\mbox{ sec}^{-1}$.
The X-ray flux increases smoothly over the first $4-5$ days of the flare, then appears to increase more steeply until the peak is reached. 

\subsection{Optical Data Reduction}

The Lowell Observatory observations were taken, reduced, and processed by the Program for Extragalactic Astronomy (PEGA) group at GSU. 
Bias/zero and flat calibration frames were taken along with the PG 1553+11 object frames.
Dark calibration frames were not required for the 1.8m telescope's CCD since the chip contains a dark pixel strip. 
All data reduction utilized standard NOAO IRAF \footnote{IRAF is distributed by the National Optical Astronomy Observatories, which are operated by the Association of the Universities for Research in Astronomy, Inc., under cooperative agreement with the National Science Foundation.}  routines including {\tt ccdproc}, {\tt flatcombine}, and {\tt zerocombine}. 
All data processing and 7 arcsec aperture photometry was done using the {\tt ccdphot} routine, written by Marc Buie, in IDL. 

The Abastumani Observatory observations were processed using the same check stars as the Lowell data. 
The frames were obtained using the Peltier cooled ST-6 CCD Camera attached to the Newtonian focus of the 70-cm meniscus telescope \citep{kur02, kur04}.
All observations were performed using combined filters of glasses which match the standard B (Johnson) and Rc (Cousins) bands.
The CCD frames were reduced using {\tt DaophotII} \citep{ste87}.
The Abastumani data were subsequently averaged into 4 point data bins in order to reduce the data's error and more clearly show any trends in the data. 

Observations were taken at the Braeside Observatory during only two nights of the campaign, but with very intensive coverage in order to observe any optical microvariability behavior present on those nights.
The 2003 April 26 data feature 70 second integration times, the April 28 data feature 120 second integrations. 
These data were also binned to reduce error and more clearly show trends. 
The 2003 April 26 data were averaged into 5 point bins, the April 28 data into 4 point bins.

\subsection{Results of Optical Observations}

The optical light curves displayed in this paper use a magnitude scale based on differential magnitudes compared to check, or calibration, stars Nos. 1 and 3. 
Check star 1 has a B magnitude of 14.42 and an R magnitude of 13.20, check star 3 features B 14.24 and R 12.95 according to the USNO-B1 survey plates, which have an error of about 0.25 magnitudes \citep{mon03}.
All differential magnitude data from Lowell and Abastumani Observatories are shown in Table 3.
According to our observed differential magnitudes and the known check star magnitude, PG 1553+11 ranges from about B = 14.54 to 14.66 and R = 13.55 to 13.66 during the RXTE campaign.
In early March preceding this campaign, observations taken at Lowell Observatory showed PG 1553+11's R magnitude at nearly 13.4, a brighter than average state for this object \citep{mil88}.
It was those observations which were subsequently used to trigger the RXTE ToO proposal for this object.

Subsequent optical observations began on 2003 April 24, and continued with regular monitoring to the end of the RXTE campaign.
The optical lightcurves from this time are displayed in Figure 1.
The R-band was sampled more frequently than the B-band, particularly during the first half of the campaign. 
Over the course of our observations, the B data show a sporadic decrease in brightness, but no behavior similar to the X-ray flare.
The R data show a mild flare of about 0.1 magnitudes in the middle of the campaign, but this $\sim$10\% increase in brightness is much smaller than the more than doubling in brightness of the X-ray flux. 
During the last half of the campaign, when there are nearly simultaneous R and B observations, the behavior in these filters is nearly identical. 

The intensively sampled R band data, shown in Table 4, from 2003 April 26 and 28 are displayed separately, in Figure 3, to emphasize any microvariability trends present on those nights. 
The lightcurves displayed in Figure 3 suggest the possibility that microvariability may be present in the days preceding the X-ray flare.
On 2003 April 26, small flares of variations at $\sim$2$\sigma$ level may be present in PG 1553+11 but are not statistically significant.
On 2003 April 28, no distinct flares are observed.
The most distinct feature on the 28th is a small but rather steady increase in brightness over an 8 hour timespan. 
Previous observations of other XBLs indicated that they do not exhibit microvariability nearly as much as RBLs \citep{mil99}.
The lack of significant microvariability observed for PG 1553+11 is consistent with variability behavior observed in other XBLs.

In Figure 2, our observations show a flare in the R band in late May and an elevated flux state on 2003 July $24 - 26$, more than two months after completion of the RXTE campaign.
The flare features a clear peak at 0.15 magnitudes brighter than the R state at the end of our RXTE campaign, which ended just after the X-ray flare appears to have peaked. 
This magnitude increase represents a flux increase of almost 15\%; a smaller flux increase than that observed for the X-ray flare, but still significant.
The later elevated R state is about 0.27 magnitudes brighter than observed in late May, corresponding to a 28\% increase in brightness and an R magnitude of about 13.47. 

\subsection{Radio Data Reduction}

The 37 GHz data were obtained with the 13.7m diameter radome-enclosed antenna of the Mets\"{a}hovi Radio Observatory in Finland.
The 37 GHz receiver is a dual horn, Dicke-switched receiver with a HEMT preamplifier and it is operated at room temperature. 
The observations are on--on observations, alternating between the source and the sky in each feed horn. 
A typical integration time to obtain one flux density data point is 1200--1600 seconds.
The signal-to-noise ratio of the data depends on the weather conditions but the general sensitivity is radome-limited so that the detection limit under optimal weather conditions is $\sim$0.2 Jy.
The data points with S/N $<$ 4 are handled as non-detections.
As a primary flux calibrator we use DR21, and as a secondary calibrator 3C 84 and 3C 274. 
The error bars in the data include the contribution from the measurement rms and the uncertainty of the absolute calibration. 
For more details about the Mets\"{a}hovi observing system and data reduction see \citet{ter98}.
 
The University of Michigan radio data were obtained using a 26m prime focus paraboloid equipped with transistor-based radiometers operating at central frequencies of 4.8, 8.0 and 14.5 GHz and room-temperature wide-band HEMPT amplifiers (with a width of $\sim$10\% of the observing frequency).
Measurements at all three frequencies utilized rotating, dual-horn polarimeter feed systems which permitted both total flux density and linear polarization to be measured. 
An on--off observing technique was used at 4.8 GHz, and an on--on technique at the other two frequencies.
A typical observation consisted of 8 to 16 individual measurements over a 25 to 45 minute period (depending on frequency) for this weak source. 
A source selected from a grid of calibrators was observed every 1 to 2 hours. 
The flux scale was set by observations of Cassiopeia A. 
Details of the calibration and analysis technique are given in \citet{all85}.

\subsection{Results of Radio Observations}

The radio data, shown in Table 5, are not as densely sampled as the X-ray data, but the combined data from 14.5 and 37.0 GHz span most of the same time window as the X-ray campaign, as displayed in Figure 1.
The flux at 14.5 GHz appears to be steady for the duration of our RXTE campaign. 
The best data, i.e. data taken from 2003 May 4 onward, feature an average flux of 0.23 $\pm$ 0.02 Jy.
One data point taken at 4.8 GHz gave a flux of 0.37 $\pm$ 0.02 Jy on May 5.
The flux at 37.0 GHz is very noisy, with a signal/noise ratio of less than 4 for many points. 
The best data have an average flux of 0.25 $\pm$ 0.04 Jy, which is near the detection limit of the telescope.
No significant changes in the flux occurred during this time.

In late July, the UMRAO data show a distinct increase in flux at 4.8 and 14.5 GHz, during which a Mets\"{a}hovi Observatory data point shows an elevated flux state at 37.0 GHz. 
The magnitude of this flux increase is about 0.1 Jy at both frequencies, representing about a 30\% increase in flux at 4.8 GHz and about a 50\% increase at 14.5 GHz.
This activity, shown in Figure 2, is most likely not related to the aforementioned X-ray and optical flares.
According to \citet{mar04}, coordinated multiwavelength variability patterns usually occur on timescales of a few weeks at the most, and the observed increase in radio flux does not occur until about two months after the optical flare.
Our optical observations at this time show a bright state in the R-band for this object; even brighter than the March observations which triggered our RXTE campaign.
While there is not a well-defined R band flare observed, this high state may be related to the increase in radio flux.

\subsection{Summary of Observations}

The X-ray flare, during which the flux more than doubles, very clearly stands out in the RXTE data.
None of the other wavebands show such a dramatic and well-defined increase in flux, however the R band data show a flare of about 0.1 magnitudes ending just before the X-ray flare begins. 
The optical brightness and 37.0 GHz flux show slight decreases as the X-ray flux increases. 
The 14.5 GHz data remains stable throughout the X-ray flare. 
In late May, the Lowell Observatory R-band observations show a distinct flare of $\sim$0.15 magnitudes, or a 15\% increase in flux.
The optical flare was of a smaller magnitude than the earlier X-ray flare, but both flares take about 10 days to reach maximum brightness, so it is possibile that the two flares are related.

\section{PG 1553+11 as an Extreme XBL}

The SEDs for the three week campaign are presented in Figure 4.
One SED was created for each five days of the RXTE campaign in order to study the evolution of the spectral shape. 
They are presented as a log $\nu$F($\nu$) vs. log($\nu$) plot with flux in units of Jy and frequency in Hz.
The optical differential magnitudes were coverted to apparent magnitudes, then to fluxes using \citet{cox01}.
An average frequency of $1.5\times10^{18}$ Hz, corresponding to about 6 keV, was used for the RXTE data.

The general shape of each SED for PG 1553+11 is consistent with that of an XBL; the emitted power is lowest at radio wavelengths and reaches a peak in the UV/soft X-ray regime.
The SEDs look very similar to an SED published by \citet{gio95} which was compiled from non-simultaneous archive data in similar wavebands. 
Recent works by, e.g., \citet{don05} and \citet{per05} confirm that this object's X-ray spectrum is that of an XBL.

Following the work of \citet{lan86}, a parabola of the form \[f(x) = -c(x-a)^2 + b\] was fit to the SED data in order to estimate the frequency of highest emitted energy, where x = log($\nu$) and f(x) = log $\nu$F($\nu$).
The best fit is plotted with each SED, with the best fit values given in Table 6.
The SED peak power values, $\nu_{max}$F($\nu_{max}$), are $7.3\mbox{ -- }8.4\times10^{-11}\mbox{ erg cm}^{-2}\mbox{ sec}^{-1}$, with values increasing as the X-ray flux increases.
These peak values occur at 86 -- 175 nm, and the wavelength at which the SED peak occurs decreases as the X-ray flux increases.
These peak power and wavelength values agree well with values found for other XBLs using parabolic fits, e.g. PKS 2155-304 \citep{gio05}, Mkn 421 \citep{mas04a}, and Mkn 501 \citep{mas04b} (see Table 7 for compared values).
As was found for PG 1553+11, as the X-ray flux increases, $\nu_{max}$F($\nu_{max}$) increases for these other XBLs. 
For Mkn 421, Mkn 501, and PG 1553+11, the SED broadens as the X-ray flux increases. 
This general agreement between PG 1553+11's behavior and that of other XBLs is important because, while many multiwavelength campaigns have been performed on the other three XBLs, no campaigns have been done using PG 1553+11. 
One surprising find was that the SED of PG 1553+11 peaks at a lower frequency than the other three XBLs. 

Following the work of \citet{per96}, the logarithmic ratios of the X-ray to radio fluxes were calculated, and presented in Table 6.
The parabolic SED fit was used to interpolate optical and radio fluxes at 250 nm and 5 GHz, respectively. 
The spectral fits from the RXTE data were used to calculate the X-ray flux from 2 -- 10 keV.
The flux ratios range from -4.37 to -3.88, increasing with the X-ray flux.
This is the most exciting result from the SED analysis; these values are consistent with PG 1553+11 being an extreme XBL, where the cutoff between RBLs and XBLs is -5.5 and extreme XBLs have a value of -4.5 or greater \citep{rec03}.
Compared to flux ratios calculated for the previously discussed XBLs (see Table 7), PG 1553+11 is significantly more extreme than Mkn 421 and Mkn 501 even in a low X-ray state, and in most states is more extreme than PKS 2155-304 \citep{per96}.

\section{Comparing PG 1553+11 to Blazar Emission Models}

Recent models of blazars address two primary questions: (1) the source of the IC photons and (2) the general properties of the jet. 
According to Synchrotron Self-Compton (SSC) models, the IC photons are synchrotron photons which have been upscattered to high energies \citep{har96}. 
According to External Radiation Compton (ERC) models, the source of IC photons is something other than the jet, for example the central accretion disk or the broad line region (BLR) \citep{har96}. 
The ERC model predicts that a blazar's X-ray flux variations will lag behind optical and radio variations \citep{gmad96}.
The SSC model predicts that the X-ray and radio fluxes will vary nearly simultaneously, provided that the X-ray emission is produced by the IC process.
Also, the SSC model predicts that UV, optical, and IR flux variations will lead variations in the radio and X-ray fluxes.
However, if the X-ray emission is produced by the synchrotron process, X-ray flaring will precede optical and radio flares.

The ERC and SSC models in which the X-ray emission is due to the IC process predict that an optical flare would precede the X-ray flare.
If this were observed, it would contradict PG 1553+11's classification as an XBL. 
A distinct increase in brightness of about 0.1 magnitudes was observed at optical wavelengths a few days before the X-ray flare was observed, but since PG 1553+11 is an XBL, this small flare is most likely unrelated to the subsequent X-ray flare.
A larger increase in R brightness was observed in 2003 March before the RXTE campaign began, but this occurred too long before the X-ray flare for the two events to be related \citep{mar04}.

Since PG 1553+11 is an XBL, the radio, optical, and X-ray emission are thought to be produced by the synchrotron process.
The X-ray emission should originate from the more compact region closer to the central engine than emission at longer wavelengths \citep{urr95}.
Thus, any flaring in the radio and optical flux should distinctly follow a corresponding flare in the X-ray flux.
The SSC model in which X-ray emission is produced by the synchrotron process is supported by the flares observed less than two weeks apart in the X-ray and then optical R wavebands. 
According to this model, a radio flare should have followed the optical flare.
However, only a few radio observations are available during this time.
The absence of observed large radio variations prevents providing strong constraints on the models.
This SSC model has been used to explain the behavior of other, more well-studied blazars, such as PKS 2155-304 \citep{cip03}.
PG 1553+11 is a relatively high reshift XBL ($z = 0.36$), and may be considered an extreme counterpart to low redshift XBLs.

\section{Conclusions}

In summary, a major X-ray flare, spanning $\sim$2 weeks, was observed for PG 1553+11 during which the X-ray flux more than tripled. 
Coordinated optical observations detected an optical flare which followed the X-ray flare by 12 days and which may be related to it.
However, due to the termination of the X-ray monitoring, the case for this relationship, while plausible, cannot be convincingly made. 
Radio observations covering this same time interval show little variability.
Optical intraday observations of PG 1553+11 found no significant evidence of microvariability.

The observations support the SSC model for blazars in which the X-ray emission is primarily due to the synchrotron process.
An investigation of the SED for PG 1553+11 is consistent with this object being an extreme XBL.
Future observations would benefit from longer monitoring programs and denser sampling in all observed wavebands so that complete flares may be well-defined in all wavebands, whether they lead or lag each other.

\section{Acknowledgments}

The authors acknowledge many helpful conversations with Paul Wiita concerning these observations and their implications. 
MAO, HRM, and KM are supported in part by the Program for Extragalactic Astronomy's Research Program Enhancement funds from Georgia State University. 
The RXTE observations were supported by NASA grant NAG5-13733.
MAO, HRM, KM, JPM, and AMC thank Lowell Observatory and Boston University for generous allocations of observing time on the Perkins telescope with the Loral camera.
The UMRAO facility is partially supported by a series of grants from the NSF and by the University of Michigan. 
OMK gratefully acknowledges the hospitality of the Astrophisikalisches Institute Potsdam, Landessternwarte Heidelberg and thanks Dr. G. M. Richter and Prof. S. Wagner for their kind collaboration of many years and invaluable financial support.

\input{tab1}
\input{tab2}
\input{tab3}
\input{tab4}
\input{tab5}
\input{tab6}
\input{tab7}

\begin{figure}
\includegraphics[angle=270, scale=0.65]{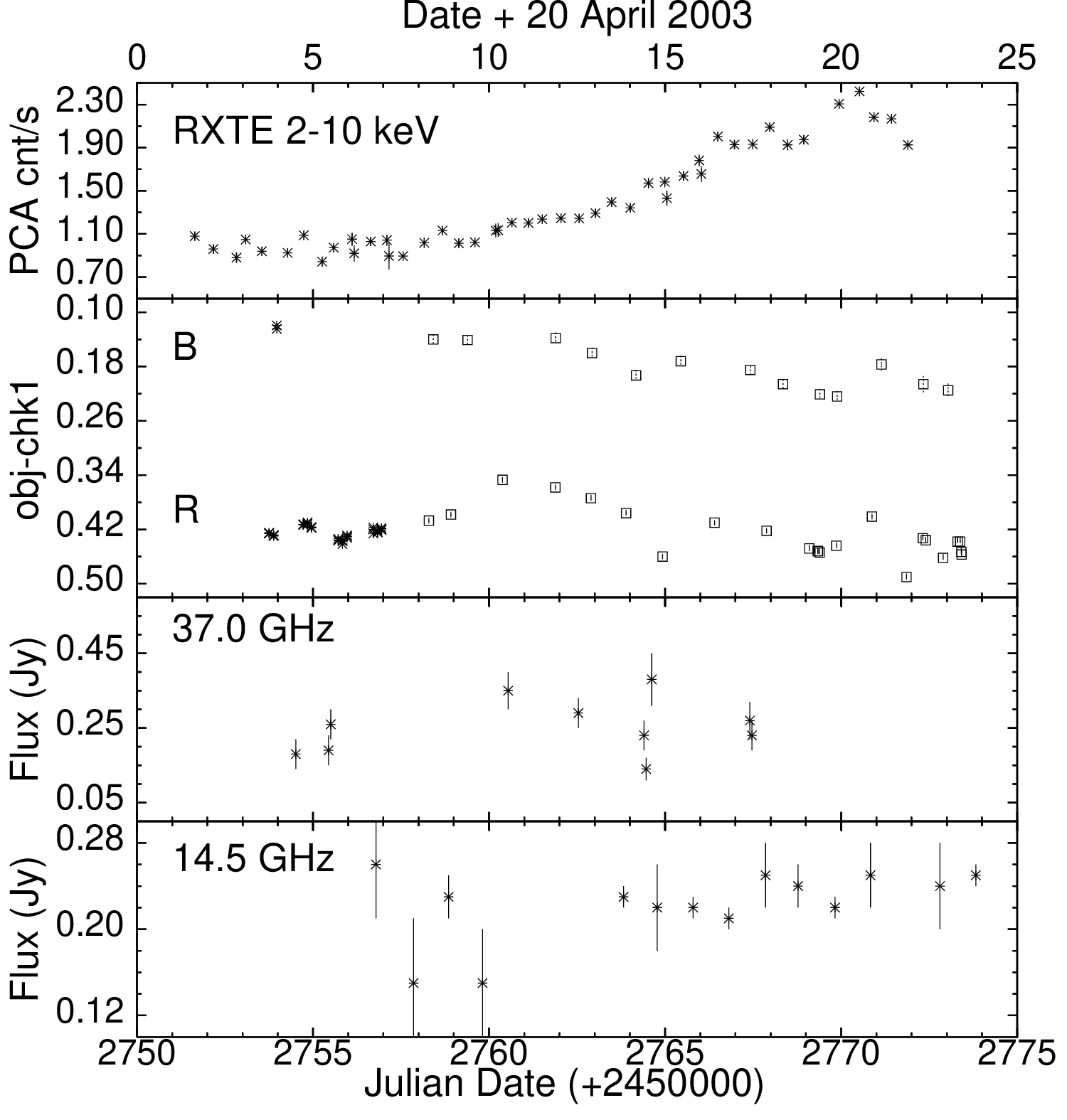}
\figcaption{All data (not including Braeside Obs) taken during the RXTE campaign. 
For R and B bands; asterisk = Lowell Obs, square = Abastumani Obs. 
For all optical and X-ray data, the error bars are smaller than the plotted points.}
\end{figure}

\begin{figure}
\includegraphics[angle=270, scale=0.65]{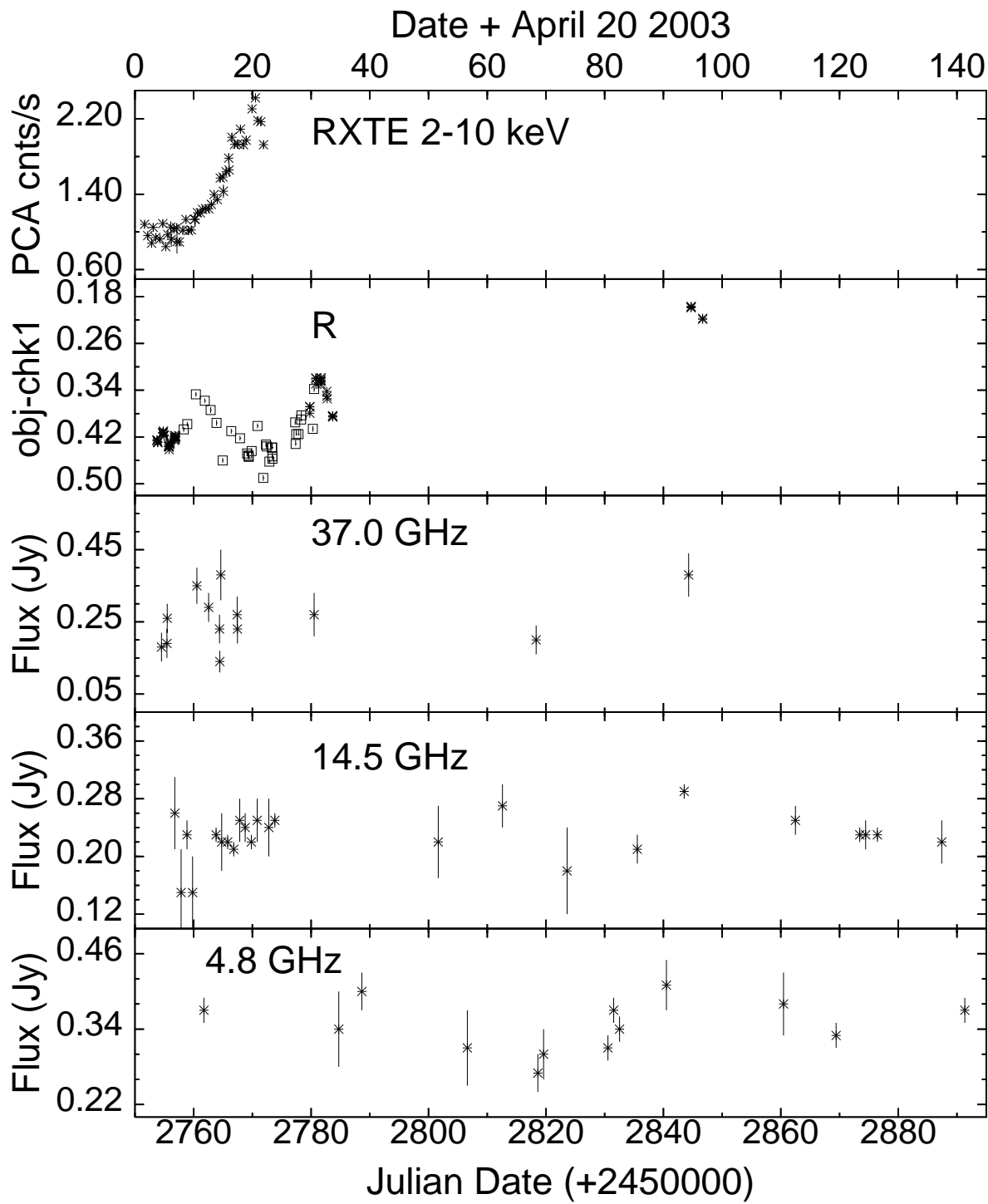}
\figcaption{All data (not including Braeside Obs R data or any B data) taken from April through September of 2003. 
For R band; asterisk = Lowell Obs, square = Abastumani Obs. 
For all optical and X-ray data, the error bars are smaller than the plotted points.}
\end{figure}

\begin{figure}
%\input{f2a.tex}
%\input{f2b.tex}
%\plottwo{f3a.eps}{f3b.eps}
\includegraphics[angle=270, scale=0.35]{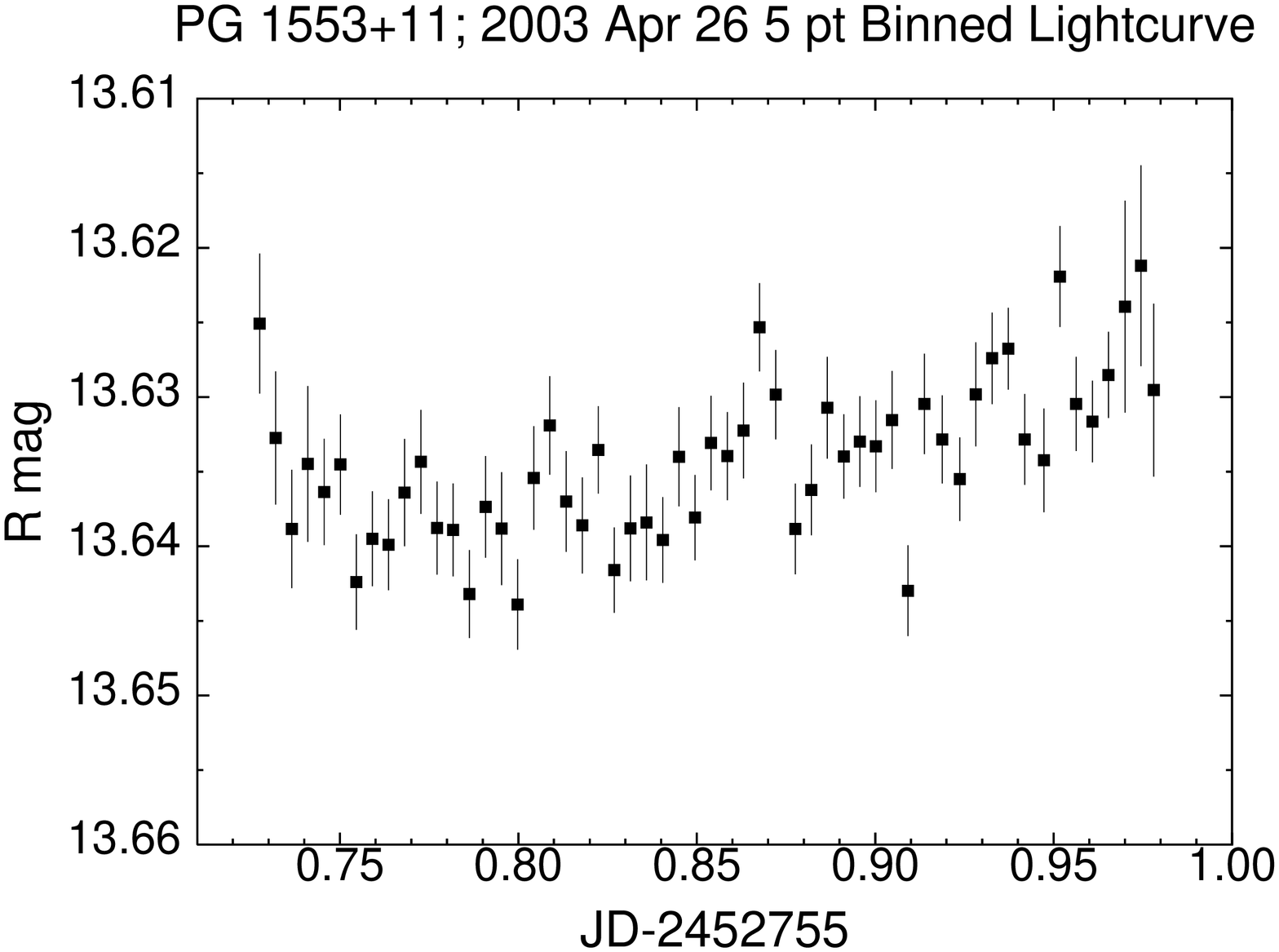}
\includegraphics[angle=270, scale=0.35]{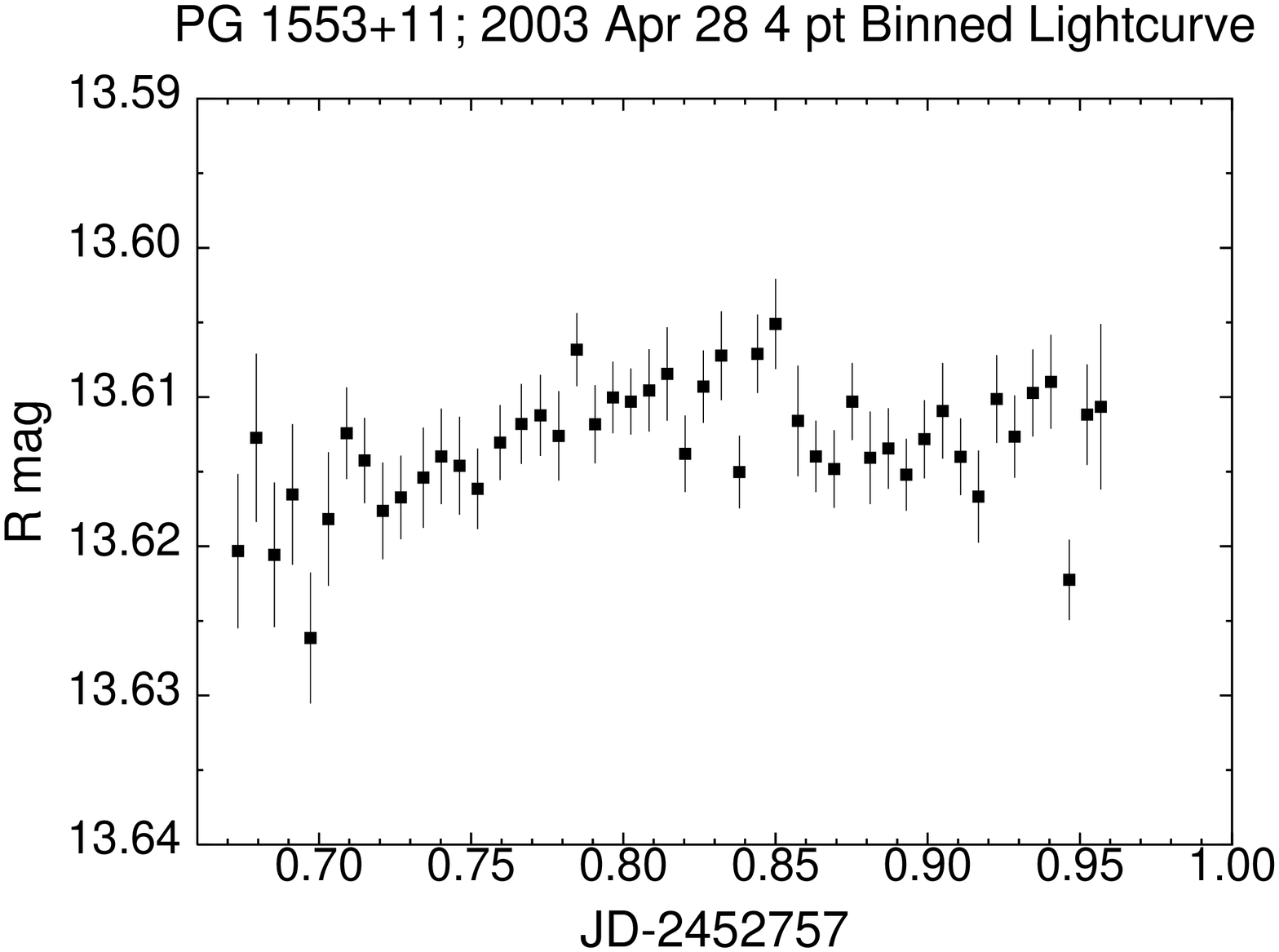}
\figcaption{Data taken at Braeside Observatory.}
\end{figure}

\begin{figure}
\includegraphics[angle=270, scale=0.65]{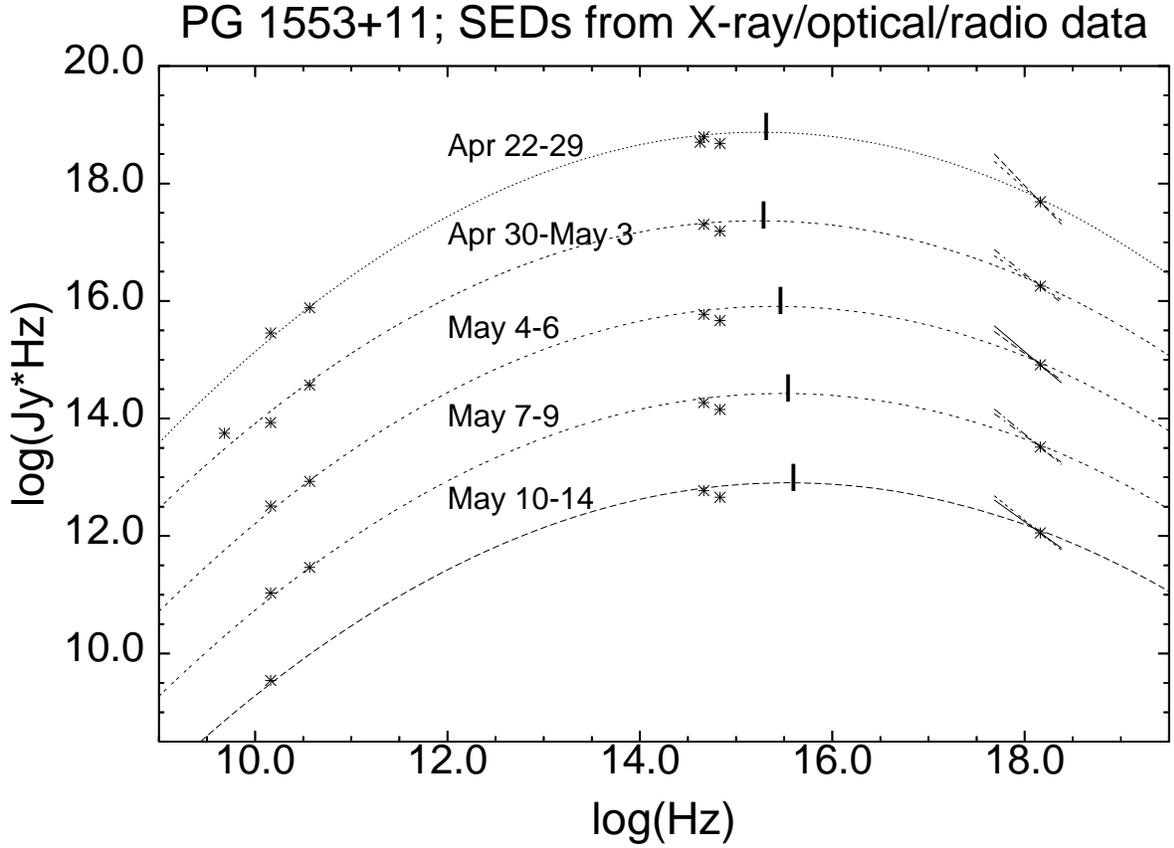}
\figcaption{Spectral Energy Distributions of all 4.8 GHz, 14.5 GHz, 37.0 GHz, Lowell \& Abastumani Observatories R \& B band, and RXTE data taken during the campaign. 
Vertical lines indicate the peak log(Jy*Hz) points for each SED. 
Short lines indicate the spectral slopes of the X-ray data. 
The y-axis gives the correct values for May $10-14$, data from other epochs are offset by increments of 1.5. }
\end{figure}

\end{document}

%% file: tab1.tex
\begin{deluxetable}{cccccc}
\tablewidth{0pt}
\tablecaption{Summary of Observations.}
\tablehead{
\colhead{Regime} & \colhead{Observatory} & \colhead{Dates of Observations} & \colhead{\# of Observations}
}
\startdata
$2-10$ keV & RXTE & 2003 April 22$ - $May 11 & 46 \\
R-band & Lowell & 2003 March $6 - 7$ & 10 \\
& & 2003 April $24 - 27$ & 33 \\
& & 2003 May $20 - 24$ & 19 \\
& & 2003 July 24 \& 26 & 5 \\
& Braeside & 2003 April 26 \& 28 & 468 \\
& Abastumani & 2003 April 28$ - $May 21 & 31 \\
B-band & Lowell & 2003 April 24 & 2 \\
& Abastumani & 2003 April 28$ - $May 20 & 17 \\
37.0 GHz & Mets\"{a}hovi & 2003 April 25$ - $May 21 & 23 \\
14.5 GHz & UMRAO & 2003 April 27$ - $Sept 4 & 24 \\
4.8 GHz & UMRAO & 2003 May 2$ - $Sept 8 & 13 \\
\enddata
\end{deluxetable}

%% file: tab2.tex
\begin{deluxetable}{cccccccc}
\tablewidth{0pt}
\tablecaption{PG 1553+11 RXTE Light Curve Data.}
\tabletypesize{\scriptsize}
\tablehead{
\colhead{UT Date} & \colhead{JD-2450000} & \colhead{2 -- 10 keV Flux\tablenotemark{a}} & \colhead{Error\tablenotemark{a}} & \colhead{Cnts/Sec} & \colhead{Error} & \colhead{Spectral Slope} & \colhead{Error}
}
\startdata
2003 April 22 &2751.64 &5.56 &0.21 &1.080 &0.040 &2.255 &0.334\\
&2752.17 &4.62 &0.19 &0.960 &0.039 & 2.281 &0.285\\
2003 April 23 &2752.83 &4.12 &0.23 &0.880 &0.048 &2.745 &0.775\\
&2753.09 &4.87 &0.18 &1.048 &0.039 &2.021 &0.387\\
2003 April 24 &2753.55 &4.95 &0.21 &0.940 &0.039 &2.569 &0.301\\
&2754.28 &4.84 &0.21 &0.924 &0.040 &2.712 &0.479\\
2003 April 25 &2754.74 &5.31 &0.20 &1.088 &0.042 &2.506 &0.402\\
&2755.26 &5.15 &0.28 &0.844 &0.046 &2.756 &0.505\\
2003 April 26 &2755.59 &4.87 &0.19 &0.972 &0.038 &2.517 &0.440\\
&2756.11 &4.21 &0.24 &1.052 &0.061 &2.091 &0.655\\
&2756.17 &3.12 &0.26 &0.920 &0.076 &3.191 &1.688\\
2003 April 27 &2756.64 &5.18 &0.20 &1.030 &0.039 &2.673 &0.410\\
&2757.10 &5.14 &0.25 &1.042 &0.051 &2.495 &0.522\\
&2757.16 &5.30 &0.74 &0.896 &0.125 &3.677 &1.337\\
2003 April 28 &2757.56 &5.03 &0.23 &0.894 &0.041 &2.434 &0.419\\
&2758.16 &5.56 &0.26 &1.018 &0.047 &2.739 &0.453\\
2003 April 29 &2758.68 &5.53 &0.19 &1.132 &0.038 &2.379 &0.366\\
&2759.14 &4.90 &0.21 &1.014 &0.043 &2.396 &0.475\\
2003 April 30 &2759.60 &4.33 &0.18 &1.022 &0.043 &2.125 &0.326\\
&2760.19 &4.72 &0.26 &1.132 &0.063 &2.236 &0.714\\
&2760.26 &5.55 &0.31 &1.136 &0.063 &1.979 &0.431\\
2003 May 1 &2760.65 &6.09 &0.19 &1.204 &0.037 &2.352 &0.291\\
&2761.12 &6.06 &0.21 &1.202 &0.041 &2.092 &0.242\\
2003 May 2 &2761.51 &6.01 &0.20 &1.238 &0.041 &2.314 &0.356\\
&2762.04 &6.40 &0.21 &1.246 &0.041 &2.030 &0.283\\
2003 May 3 &2762.56 &6.26 &0.21 &1.244 &0.041 &2.447 &0.339\\
&2763.02 &5.83 &0.18 &1.292 &0.039 &2.158 &0.264\\
&2763.48 &5.86 &0.18 &1.396 &0.044 &2.156 &0.291\\
2003 May 4 &2764.01 &6.90 &0.21 &1.342 &0.040 &2.266 &0.253\\
2003 May 5 &2764.53 &7.71 &0.24 &1.572 &0.048 &2.383 &0.280\\
&2764.99 &7.94 &0.25 &1.582 &0.049 &2.054 &0.285\\
&2765.05 &7.65 &0.37 &1.432 &0.070 &2.276 &0.345\\
2003 May 6 &2765.52 &8.22 &0.22 &1.638 &0.043 &2.359 &0.216\\
&2765.97 &8.68 &0.27 &1.782 &0.055 &2.275 &0.301\\
&2766.03 &9.56 &0.42 &1.656 &0.073 &2.445 &0.381\\
2003 May 7 &2766.50 &9.09 &0.21 &2.004 &0.046 &2.273 &0.273\\
&2766.97 &10.3 &0.22 &1.928 &0.041 &2.310 &0.215\\
&2767.49 &10.3 &0.25 &1.932 &0.048 &2.284 &0.136\\
2003 May 8 &2767.97 &11.0 &0.20 &2.090 &0.038 &2.323 &0.187\\
&2768.48 &9.59 &0.25 &1.926 &0.051 &2.131 &0.182\\
2003 May 9 &2768.94 &10.6 &0.23 &1.974 &0.043 &2.308 &0.195\\
2003 May 10 &2769.94 &12.0 &0.22 &2.306 &0.043 &2.259 &0.172\\
2003 May 11 &2770.52 &12.0 &0.22 &2.422 &0.044 &2.190 &0.169\\
&2770.93 &12.0 &0.24 &2.182 &0.044 &2.460 &0.182\\
&2771.43 &11.0 &0.23 &2.168 &0.045 &2.194 &0.142\\
2003 May 12 &2771.90 &9.76 &0.20 &1.926 &0.040 &2.130 &0.163\\
\enddata
\tablenotetext{a}{In units of $10^{-12}\mbox{ erg cm}^{-2}\mbox{ sec}^{-1}$.}
\end{deluxetable}

%% file: tab3.tex
\begin{deluxetable}{cccccccccc}
\tablewidth{0pt}
\tablecaption{PG 1553+11 R- and B-band Light Curve Data.}
\tabletypesize{\scriptsize}
\tablehead{
\colhead{UT Date} & \colhead{JD-2450000} & \colhead{Filter} & \colhead{Obj-Chk1\tablenotemark{a}} & \colhead{Error} & \colhead{Obj-Chk3} & \colhead{Error} & \colhead{Chk1-Chk3} & \colhead{Error} & \colhead{Observatory\tablenotemark{b}} }
\startdata
2003 April 24 &2753.972 &B &0.120 &0.004 &0.168 &0.004 &0.048 &0.004 &LO\\
&2753.974 &B &0.125 &0.004 &0.169 &0.004 &0.045 &0.004 &LO\\

2003 April 28 &2758.417 &B &0.140 &0.007 &0.174 &0.006 &0.034 &0.006&AB\\
2003 April 29 &2759.390 &B &0.141 &0.007 &0.127 &0.007 &-0.014 &0.007 &AB\\
2003 May 2 &2761.892 &B &0.138 &0.008 &0.105 &0.008 &-0.033 &0.008 &AB\\
2003 May 3 &2762.923 &B &0.160 &0.008 &0.147 &0.008 &-0.013 &0.007 &AB\\
2003 May 4 &2764.174 &B &0.193 &0.008 &0.230 &0.008 &0.037 &0.007 &AB\\
2003 May 5 &2765.448 &B &0.172 &0.008 &0.179 &0.008 &0.007 &0.007 &AB\\
2003 May 7 &2767.418 &B &0.185 &0.006 &0.212 &0.006 &0.027 &0.006 &AB\\
2003 May 8 &2768.350 &B &0.206 &0.007 &0.231 &0.007 &0.026 &0.006 &AB\\
2003 May 9 &2769.400 &B &0.221 &0.008 &0.250 &0.008 &0.030 &0.007 &AB\\
2003 May 10 &2769.886 &B &0.224 &0.008 &0.257 &0.008 &0.033 &0.007 &AB\\
2003 May 11 &2771.146 &B &0.177 &0.009 &0.169 &0.009 &-0.008 &0.008 &AB\\
2003 May 12 &2772.334 &B &0.206 &0.012 &0.207 &0.011 &0.001 &0.011 &AB\\
2003 May 13 &2773.037 &B &0.215 &0.010 &0.190 &0.010 &-0.025 &0.009 &AB\\
2003 May 17 &2777.282 &B &0.163 &0.010 &0.182 &0.010 &0.019 &0.010 &AB\\
2003 May 18 &2777.822 &B &0.182 &0.011 &0.186 &0.011 &0.004 &0.010 &AB\\
&2778.355 &B &0.153 &0.014 &0.128 &0.014 &-0.025 &0.012 &AB\\
2003 May 20 &2780.436 &B &0.080 &0.010 &0.100 &0.009 &0.020 &0.009 &AB\\

2003 March 06 &2704.986 &R &0.234 &0.004 &0.359 &0.004 &0.124 &0.004 &LO\\
&2704.987 &R &0.235 &0.004 &0.359 &0.004 &0.125 &0.004 &LO\\
&2705.002 &R &0.237 &0.004 &0.359 &0.004 &0.122 &0.004 &LO\\
&2705.003 &R &0.238 &0.004 &0.357 &0.004 &0.119 &0.004 &LO\\
2003 March 07 &2706.000 &R &0.210 &0.004 &0.327 &0.004 &0.117 &0.004 &LO\\
&2706.002 &R &0.211 &0.004 &0.330 &0.004 &0.118 &0.004 &LO\\
&2706.003 &R &0.210 &0.004 &0.328 &0.004 &0.118 &0.004 &LO\\
&2706.017 &R &0.212 &0.004 &0.330 &0.004 &0.117 &0.004 &LO\\
&2706.018 &R &0.212 &0.004 &0.329 &0.004 &0.117 &0.004 &LO\\
&2706.020 &R &0.211 &0.004 &0.329 &0.004 &0.118 &0.004 &LO\\
2003 April 24 &2753.750 &R &0.426 &0.004 &0.552 &0.004 &0.126 &0.004 &LO\\
&2753.752 &R &0.426 &0.004 &0.547 &0.004 &0.121 &0.004 &LO\\
&2753.754 &R &0.425 &0.004 &0.550 &0.004 &0.125 &0.004 &LO\\
&2753.880 &R &0.430 &0.004 &0.554 &0.004 &0.125 &0.004 &LO\\
&2753.882 &R &0.429 &0.004 &0.553 &0.004 &0.124 &0.004 &LO\\
&2753.884 &R &0.428 &0.004 &0.554 &0.004 &0.126 &0.004 &LO\\
2003 April 25 &2754.723 &R &0.413 &0.004 &0.537 &0.004 &0.125 &0.004 &LO\\
&2754.725 &R &0.413 &0.004 &0.536 &0.004 &0.124 &0.004 &LO\\
&2754.728 &R &0.413 &0.004 &0.540 &0.004 &0.127 &0.004 &LO\\
&2754.840 &R &0.413 &0.004 &0.540 &0.004 &0.127 &0.004 &LO\\
&2754.842 &R &0.412 &0.004 &0.540 &0.004 &0.128 &0.004 &LO\\
&2754.844 &R &0.410 &0.004 &0.541 &0.004 &0.131 &0.004 &LO\\
&2754.955 &R &0.417 &0.004 &0.540 &0.004 &0.123 &0.004 &LO\\
&2754.957 &R &0.418 &0.004 &0.543 &0.004 &0.125 &0.004 &LO\\
&2754.959 &R &0.417 &0.004 &0.541 &0.004 &0.124 &0.004 &LO\\
2003 April 26 &2755.714 &R &0.434 &0.004 &0.560 &0.004 &0.127 &0.004 &LO\\
&2755.716 &R &0.436 &0.004 &0.561 &0.004 &0.125 &0.004 &LO\\
&2755.718 &R &0.436 &0.004 &0.562 &0.004 &0.125 &0.004 &LO\\
&2755.836 &R &0.437 &0.004 &0.562 &0.004 &0.125 &0.004 &LO\\
&2755.838 &R &0.442 &0.004 &0.565 &0.004 &0.124 &0.004 &LO\\
&2755.840 &R &0.437 &0.004 &0.562 &0.004 &0.126 &0.004 &LO\\
&2755.968 &R &0.433 &0.004 &0.564 &0.004 &0.131 &0.004 &LO\\
&2755.970 &R &0.429 &0.004 &0.561 &0.004 &0.132 &0.004 &LO\\
&2755.972 &R &0.431 &0.004 &0.565 &0.004 &0.134 &0.004 &LO\\
2003 April 27 &2756.711 &R &0.426 &0.004 &0.549 &0.004 &0.123 &0.004 &LO\\
&2756.713 &R &0.417 &0.004 &0.545 &0.004 &0.128 &0.004 &LO\\
&2756.715 &R &0.420 &0.004 &0.548 &0.004 &0.128 &0.004 &LO\\
&2756.833 &R &0.422 &0.004 &0.548 &0.004 &0.126 &0.004 &LO\\
&2756.835 &R &0.424 &0.004 &0.551 &0.004 &0.127 &0.004 &LO\\
&2756.837 &R &0.424 &0.004 &0.551 &0.004 &0.127 &0.004 &LO\\
&2756.942 &R &0.418 &0.004 &0.541 &0.004 &0.123 &0.004 &LO\\
&2756.944 &R &0.421 &0.004 &0.543 &0.004 &0.122 &0.004 &LO\\
&2756.947 &R &0.419 &0.004 &0.542 &0.004 &0.123 &0.004 &LO\\

2003 April 28 &2758.293 &R &0.407 &0.004 &0.516 &0.004 &0.109 &0.003 &AB\\
2003 April 29 &2758.917 &R &0.398 &0.004 &0.502 &0.004 &0.105 &0.003 &AB\\
2003 April 30 &2760.385 &R &0.347 &0.004 &0.449 &0.004 &0.102 &0.003 &AB\\
2003 May 2 &2761.885 &R &0.358 &0.004 &0.462 &0.004 &0.105 &0.003 &AB\\
2003 May 3 &2762.897 &R &0.374 &0.005 &0.483 &0.004 &0.108 &0.004 &AB\\
2003 May 4 &2763.895 &R &0.396 &0.005 &0.508 &0.004 &0.112 &0.004 &AB\\
2003 May 5 &2764.930 &R &0.460 &0.004 &0.554 &0.004 &0.095 &0.003 &AB\\
2003 May 7 &2766.406 &R &0.410 &0.004 &0.523 &0.004 &0.113 &0.003 &AB\\
2003 May 8 &2767.885 &R &0.422 &0.004 &0.542 &0.004 &0.120 &0.003 &AB\\
2003 May 9 &2769.096 &R &0.448 &0.005 &0.568 &0.004 &0.120 &0.004 &AB\\
&2769.333 &R &0.452 &0.005 &0.572 &0.005 &0.120 &0.004 &AB\\
&2769.345 &R &0.452 &0.005 &0.573 &0.005 &0.121 &0.004 &AB\\
&2769.390 &R &0.454 &0.006 &0.573 &0.005 &0.119 &0.004 &AB\\
2003 May 10 &2769.864 &R &0.444 &0.005 &0.566 &0.005 &0.123 &0.004 &AB\\
2003 May 11 &2770.876 &R &0.401 &0.004 &0.524 &0.004 &0.123 &0.004 &AB\\
2003 May 12 &2771.854 &R &0.490 &0.005 &0.575 &0.005 &0.086 &0.004 &AB\\
&2772.318 &R &0.433 &0.006 &0.537 &0.006 &0.103 &0.005 &AB\\
&2772.404 &R &0.436 &0.005 &0.546 &0.005 &0.110 &0.004 &AB\\
2003 May 13 &2772.895 &R &0.462 &0.006 &0.551 &0.006 &0.089 &0.005 &AB\\
&2773.309 &R &0.438 &0.007 &0.529 &0.007 &0.092 &0.006 &AB\\
&2773.381 &R &0.438 &0.007 &0.555 &0.007 &0.117 &0.005 &AB\\
&2773.412 &R &0.453 &0.006 &0.550 &0.006 &0.097 &0.005 &AB\\
&2773.422 &R &0.457 &0.006 &0.564 &0.006 &0.107 &0.004 &AB\\
2003 May 17 &2777.289 &R &0.395 &0.006 &0.531 &0.007 &0.136 &0.005 &AB\\
&2777.374 &R &0.432 &0.007 &0.537 &0.007 &0.105 &0.005 &AB\\
&2777.450 &R &0.415 &0.007 &0.515 &0.007 &0.099 &0.005 &AB\\
2003 May 18 &2777.855 &R &0.415 &0.007 &0.511 &0.007 &0.096 &0.006 &AB\\
&2778.298 &R &0.390 &0.006 &0.490 &0.006 &0.100 &0.005 &AB\\
&2778.342 &R &0.383 &0.007 &0.490 &0.007 &0.107 &0.006 &AB\\

2003 May 20 &2779.772 &R &0.379 &0.004 &0.502 &0.004 &0.123 &0.004 &LO\\
&2779.773 &R &0.369 &0.004 &0.494 &0.004 &0.125 &0.004 &LO\\
&2779.776 &R &0.368 &0.004 &0.492 &0.004 &0.124 &0.004 &LO\\

2003 May 21 &2780.303 &R &0.406 &0.005 &0.502 &0.005 &0.096 &0.003 &AB\\
&2780.505 &R &0.338 &0.005 &0.468 &0.005 &0.131 &0.004 &AB\\

&2780.813 &R &0.320 &0.004 &0.447 &0.004 &0.127 &0.004 &LO\\
&2780.814 &R &0.319 &0.004 &0.446 &0.004 &0.127 &0.004 &LO\\
&2780.814 &R &0.329 &0.004 &0.452 &0.004 &0.123 &0.004 &LO\\
2003 May 22 &2781.670 &R &0.323 &0.004 &0.442 &0.004 &0.120 &0.004 &LO\\
&2781.671 &R &0.320 &0.004 &0.437 &0.004 &0.118 &0.004 &LO\\
&2781.673 &R &0.330 &0.004 &0.452 &0.004 &0.122 &0.004 &LO\\
&2781.674 &R &0.326 &0.004 &0.448 &0.004 &0.122 &0.004 &LO\\
&2781.676 &R &0.323 &0.004 &0.452 &0.004 &0.130 &0.004 &LO\\
&2781.679 &R &0.319 &0.004 &0.444 &0.004 &0.125 &0.004 &LO\\
2003 May 23 &2782.670 &R &0.342 &0.004 &0.463 &0.004 &0.121 &0.004 &LO\\
&2782.686 &R &0.350 &0.004 &0.472 &0.004 &0.122 &0.004 &LO\\
&2782.733 &R &0.355 &0.004 &0.477 &0.004 &0.122 &0.004 &LO\\
2003 May 24 &2783.672 &R &0.384 &0.004 &0.505 &0.004 &0.122 &0.004 &LO\\
&2783.673 &R &0.385 &0.004 &0.509 &0.004 &0.125 &0.004 &LO\\
&2783.676 &R &0.384 &0.004 &0.509 &0.004 &0.126 &0.004 &LO\\
&2783.679 &R &0.386 &0.004 &0.509 &0.004 &0.123 &0.004 &LO\\
2003 July 24 &2844.722 &R &0.197 &0.004 &0.323 &0.004 &0.127 &0.004 &LO\\
&2844.725 &R &0.199 &0.004 &0.326 &0.004 &0.126 &0.004 &LO\\
&2844.727 &R &0.198 &0.004 &0.328 &0.004 &0.130 &0.004 &LO\\
2003 July 26 &2846.681 &R &0.219 &0.004 &0.351 &0.004 &0.132 &0.004 &LO\\
&2846.685 &R &0.217 &0.004 &0.347 &0.004 &0.130 &0.004 &LO\\

\enddata
\tablenotetext{a}{ Differential magnitudes of PG 1553+11 - check 1. Similar notation is used for the differences between PG 1553+11 and check 3, and check 1 and check 3.}
\tablenotetext{b}{ AB = Abastumani Observatory, 4-point binned data. LO = Lowell Observatory.}
\end{deluxetable}

%% file: tab4.tex
\begin{deluxetable}{cccccccc}
\tablewidth{0pt}
\tablecaption{PG 1553+11 R-band Intranight Variability Data.}
\tabletypesize{\scriptsize}
\tablehead{
\colhead{UT Date} & \colhead{JD-2450000} & \colhead{Mag\tablenotemark{a}} & \colhead{Error\tablenotemark{b}} & \colhead{UT Date} & \colhead{JD-2450000} & \colhead{Mag\tablenotemark{a}} & \colhead{Error\tablenotemark{b}}}
\startdata

2003 April 26 &2755.728 &13.625 &0.005 &2003 April 28 &2757.673 &13.620 &0.005\\
&2755.732 &13.633 &0.005 & &2757.679 &13.613 &0.006\\
&2755.737 &13.639 &0.004 & &2757.685 &13.621 &0.005\\
&2755.741 &13.635 &0.005 & &2757.691 &13.616 &0.005\\
&2755.746 &13.636 &0.004 & &2757.697 &13.626 &0.004\\
&2755.750 &13.635 &0.003 & &2757.703 &13.618 &0.005\\
&2755.755 &13.642 &0.003 & &2757.709 &13.612 &0.003\\
&2755.759 &13.640 &0.003 & &2757.715 &13.614 &0.003\\
&2755.764 &13.640 &0.003 & &2757.721 &13.618 &0.003\\
&2755.768 &13.636 &0.004 & &2757.727 &13.617 &0.003\\
&2755.773 &13.634 &0.004 & &2757.734 &13.615 &0.003\\
&2755.777 &13.639 &0.003 & &2757.740 &13.614 &0.003\\
&2755.782 &13.639 &0.003 & &2757.746 &13.615 &0.003\\
&2755.786 &13.643 &0.003 & &2757.752 &13.616 &0.003\\
&2755.791 &13.637 &0.003 & &2757.760 &13.613 &0.003\\
&2755.795 &13.639 &0.004 & &2757.767 &13.612 &0.003\\
&2755.800 &13.644 &0.003 & &2757.773 &13.611 &0.003\\
&2755.804 &13.635 &0.004 & &2757.779 &13.617 &0.003\\
&2755.809 &13.632 &0.003 & &2757.785 &13.607 &0.003\\
&2755.813 &13.637 &0.003 & &2757.791 &13.612 &0.003\\
&2755.818 &13.639 &0.003 & &2757.797 &13.610 &0.002\\
&2755.822 &13.634 &0.003 & &2757.803 &13.610 &0.002\\
&2755.827 &13.642 &0.003 & &2757.809 &13.610 &0.003\\
&2755.831 &13.639 &0.004 & &2757.814 &13.609 &0.003\\
&2755.836 &13.638 &0.004 & &2757.820 &13.614 &0.003\\
&2755.841 &13.640 &0.003 & &2757.826 &13.609 &0.002\\
&2755.845 &13.634 &0.003 & &2757.832 &13.607 &0.003\\
&2755.850 &13.638 &0.003 & &2757.838 &13.615 &0.002\\
&2755.854 &13.633 &0.003 & &2757.844 &13.607 &0.003\\
&2755.859 &13.634 &0.003 & &2757.850 &13.605 &0.003\\
&2755.863 &13.632 &0.003 & &2757.857 &13.612 &0.004\\
&2755.868 &13.625 &0.003 & &2757.863 &13.614 &0.002\\
&2755.872 &13.630 &0.003 & &2757.869 &13.615 &0.003\\
&2755.878 &13.639 &0.003 & &2757.875 &13.610 &0.003\\
&2755.882 &13.636 &0.003 & &2757.881 &13.614 &0.003\\
&2755.887 &13.631 &0.003 & &2757.887 &13.614 &0.003\\
&2755.891 &13.634 &0.003 & &2757.893 &13.615 &0.002\\
&2755.896 &13.633 &0.003 & &2757.899 &13.613 &0.003\\
&2755.900 &13.633 &0.003 & &2757.905 &13.611 &0.003\\
&2755.905 &13.632 &0.003 & &2757.911 &13.614 &0.003\\
&2755.909 &13.643 &0.003 & &2757.917 &13.617 &0.003\\
&2755.914 &13.631 &0.003 & &2757.923 &13.610 &0.003\\
&2755.919 &13.633 &0.003 & &2757.929 &13.613 &0.003\\
&2755.924 &13.636 &0.003 & &2757.935 &13.610 &0.003\\
&2755.928 &13.630 &0.004 & &2757.941 &13.609 &0.003\\
&2755.933 &13.627 &0.003 & &2757.947 &13.622 &0.003\\
&2755.937 &13.627 &0.003 & &2757.952 &13.611 &0.003\\
&2755.942 &13.633 &0.003 & &2757.957 &13.611 &0.006\\
&2755.947 &13.634 &0.004\\
&2755.952 &13.622 &0.003\\
&2755.956 &13.631 &0.003\\
&2755.961 &13.632 &0.003\\
&2755.965 &13.629 &0.003\\
&2755.970 &13.624 &0.007\\
&2755.975 &13.621 &0.007\\
&2755.978 &13.630 &0.006\\

\enddata
\tablenotetext{a}{Apparent magnitudes based on observed differential magnitudes and the known magnitude of check 1.}
\tablenotetext{b}{Data has been averaged into 5 \& 4 point bins on Apr 26 \& 28, respectively.}
\end{deluxetable}

%% file: tab5.tex
\begin{deluxetable}{cccccc}
\tablewidth{0pt}
\tablecaption{PG 1553+11 Radio Flux Data}
\tabletypesize{\scriptsize}
\tablehead{
\colhead{Frequency (GHz)} & \colhead{UT Date} & \colhead{JD-2450000} & \colhead{Flux (Jy)} & \colhead{Error} & \colhead{Observatory} }
\startdata
37.0 &2003 April 25 &2754.510 &0.18 &0.04 &Mets\"{a}hovi\\
& &2755.441 &0.19 &0.04\\
&2003 April 26 &2755.503 &0.26 &0.04\\
&2003 May 1 &2760.542 &0.35 &0.05\\
&2003 May 3 &2762.538 &0.29 &0.04\\
&2003 May 4 &2764.403 &0.23 &0.04\\
& &2764.462 &0.14 &0.03\\
&2003 May 5 &2764.622 &0.38 &0.07\\
&2003 May 7 &2767.406 &0.27 &0.05\\
& &2767.465 &0.23 &0.04\\
&2003 May 21 &2780.507 &0.27 &0.06\\
&2003 June 27 &2818.326 &0.20 &0.04\\
&2003 July 23 &2844.309 &0.38 &0.06\\
14.5 &2003 April 27 &2756.787 &0.26 &0.05 &UMRAO\\
&2003 April 28 &2757.858 &0.15 &0.06\\
&2003 April 29 &2758.849 &0.23 &0.02\\
&2003 April 30 &2759.813 &0.15 &0.05\\
&2003 May 4 &2763.822 &0.23 &0.01\\
&2003 May 5 &2764.778 &0.22 &0.04\\
&2003 May 6 &2765.796 &0.22 &0.01\\
&2003 May 7 &2766.809 &0.21 &0.01\\
&2003 May 8 &2767.854 &0.25 &0.03\\
&2003 May 9 &2768.775 &0.24 &0.02\\
&2003 May 10 &2769.829 &0.22 &0.01\\
&2003 May 11 &2770.834 &0.25 &0.03\\
&2003 May 13 &2772.806 &0.24 &0.04\\
&2003 May 14 &2773.826 &0.25 &0.01\\
&2003 June 11 &2801.661 &0.22 &0.05\\
&2003 June 22 &2812.600 &0.27 &0.03\\
&2003 July 3 &2823.587 &0.18 &0.06\\
&2003 July 15 &2835.554 &0.21 &0.02\\
&2003 July 23 &2843.531 &0.29 &0.01\\
&2003 August 10 &2862.495 &0.25 &0.02\\
&2003 August 21 &2873.453 &0.23 &0.01\\
&2003 August 22 &2874.435 &0.23 &0.02\\
&2003 August 24 &2876.450 &0.23 &0.01\\
&2003 September 4 &2887.397 &0.22 &0.03\\
4.8 &2003 May 2 &2761.740 &0.37 &0.02 &UMRAO\\
&2003 May 25 &2784.708 &0.34 &0.06\\
&2003 May 29 &2788.646 &0.40 &0.03\\
&2003 June 16 &2806.597 &0.31 &0.06\\
&2003 June 28 &2818.610 &0.27 &0.03\\
&2003 June 29 &2819.593 &0.30 &0.04\\
&2003 July 10 &2830.544 &0.31 &0.02\\
&2003 July 11 &2831.529 &0.37 &0.02\\
&2003 July 12 &2832.544 &0.34 &0.02\\
&2003 July 20 &2840.525 &0.41 &0.04\\
&2003 August 8 &2860.449 &0.38 &0.05\\
&2003 August 17 &2869.411 &0.33 &0.02\\
&2003 September 8 &2891.345 &0.37 &0.02\\
\enddata
\end{deluxetable}

%% file: tab6.tex
\begin{deluxetable}{ccccccccc}
\tablewidth{0pt}
\rotate
\tablecaption{PG 1553+11 SED fit to Parabolic Function.}
\tablehead{
\colhead{Period} & \colhead{a\tablenotemark{a}} & \colhead{b\tablenotemark{b}} & \colhead{c\tablenotemark{c}} & \colhead{$\lambda_{max}$(nm)} & \colhead{$\nu_{max}F(\nu_{max})$\tablenotemark{d}} & \colhead{$\alpha_{OX}$\tablenotemark{e}} & \colhead{$\alpha_{RO}$\tablenotemark{e}} & \colhead{log($F_{x}/F_{r}$)\tablenotemark{e}} }
\startdata
Apr $22 - 29$ &15.26$\pm$0.09 &12.87$\pm$0.07 &0.1350 &164.7 &7.47 &1.16 &0.23 &-3.94\\
Apr 30 $-$ May 3 &15.23$\pm$0.07 &12.87$\pm$0.06 &0.1260 &175.4 &7.33 &1.21 &0.28 &-4.37\\
May $4 - 6$ &15.41$\pm$0.08 &12.91$\pm$0.06 &0.1265 &117.3 &8.10 &1.14 &0.24 &-3.94\\
May $7 - 9$ &15.49$\pm$0.09 &12.93$\pm$0.07 &0.1225 &97.66 &8.43 &1.11 &0.24 &-3.88\\
May $10 - 14$ &15.54$\pm$0.12 &12.91$\pm$0.08 &0.1180 &85.93 &8.05 &1.09 &0.26 &-3.91\\
\enddata
\tablenotetext{a}{The value of log(Hz) where the peak energy, measured as log(Jy*Hz), occurs.}
\tablenotetext{b}{The peak value of log(Jy*Hz).}
\tablenotetext{c}{This value measures the curvature of the parabola.}
\tablenotetext{d}{In units of $10^{-11}\mbox{ erg cm}^{-2}\mbox{ sec}^{-1}$.}
\tablenotetext{e}{Defined following \citet{per96}: radio flux interpolated at 5 GHz and optical at 250 nm from the parabolic fit. X-ray flux at 2 keV derived from RXTE spectral slope data.}
\end{deluxetable}

%% file: tab7.tex
\begin{deluxetable}{ccccc}
\tablewidth{0pt}
\tablecaption{SED Results for Selected XBLs.}
\tablehead{
\colhead{Object} & \colhead{$\lambda_{max}$(nm)} & \colhead{$\nu_{max}F(\nu_{max})$\tablenotemark{a}} & \colhead{log($F_{x}/F_{r}$)\tablenotemark{b}} & z }
\startdata
PG 1553+11 &$86-175$ &$7.33-8.43$ &-4.37 -- -3.88 &0.36\\
PKS 2155-304 &$4-62$\tablenotemark{c} &$\sim$10\tablenotemark{c} &-4.33 &0.116\tablenotemark{d}\\
Mkn 501 &$0.01-2.5$\tablenotemark{e} &$3.15-60.0$\tablenotemark{e} &-5.16 &0.0337\tablenotemark{f}\\
Mkn 421 &$1.5-4$\tablenotemark{g} &$10.7-30.4$\tablenotemark{g} &-4.86 &0.030\tablenotemark{f}\\
\enddata
\tablenotetext{a}{In units of $10^{-11}\mbox{ erg cm}^{-2}\mbox{ sec}^{-1}$.}
\tablenotetext{b}{Values for PG 1553+11 defined following \citet{per96}: radio flux interpolated at 5 GHz, optical at 250 nm, X-ray at 2 keV. Other values taken from \citet{per96}.}
\tablenotetext{c}{\citet{gio05}.}
\tablenotetext{d}{\citet{fal93}.}
\tablenotetext{e}{\citet{mas04b}.}
\tablenotetext{f}{\citet{ulr75}.}
\tablenotetext{g}{\citet{mas04a}.}
\end{deluxetable}